\documentstyle[pra,aps,psfig]{revtex}
\newcommand{\beq}{\begin{equation}}
\newcommand{\eeq}{\end{equation}}
\newcommand{\beqa}{\begin{eqnarray}}
\newcommand{\eeqa}{\end{eqnarray}}
\newcommand{\ba}{\begin{array}}
\newcommand{\ea}{\end{array}}

\begin{document}
\draft

\twocolumn[\hsize\textwidth\columnwidth\hsize\csname
@twocolumnfalse\endcsname

\widetext 
\title{Transition from 3D to 1D in Bose Gases at Zero Temperature} 
\author{L. Salasnich$^{1}$, A. Parola$^{2}$ and L. Reatto$^{1}$} 
\address{$^{1}$Dipartimento di Fisica and INFM, Universit\`a di Milano, \\
Via Celoria 16, 20133 Milano, Italy\\
$^{2}$Dipartimento di Scienze Fisiche and INFM, 
Universit\`a dell'Insubria, \\
Via Valeggio 11, 23100 Como, Italy}

\maketitle

\begin{abstract} 
We investigate the effects of dimensional reduction 
in Bose gases induced by a strong harmonic confinement 
in the transverse cylindric radial direction. 
By using a generalized Lieb-Liniger theory, based 
on a variational treatment of the transverse 
width of the Bose gas, we analyze the transition 
from a 3D Bose-Einstein condensate to the 1D Tonks-Girardeau gas. 
The sound velocity and the frequency of the lowest 
compressional mode give a clear signature of the 
regime involved. We study also the case of 
negative scattering length 
deriving the phase diagram of the Bose gas 
(uniform, single soliton, multi soliton and collapsed) 
in toroidal confinement. 
\end{abstract}


]

\narrowtext 

\section{Introduction} 

Recently the 1D regime has been experimentally 
achieved for dilute Bose gases in optical traps 
at ultra-low temperatures 
by imposing a strong cylindric radial confinement,  
which inhibits transverse excitations [1]. 
While a 3D dilute Bose gas is well described by the 
Gross-Pitaevskii (GP) theory [2], in the 1D regime 
quantum fluctuations become important 
and an appropriate theoretical description requires 
the many-body approach introduced by Lieb and Liniger 
many years ago [3] and recently extended by 
Lieb, Seiringer and Yngvason [4]. 
In particular, at sufficiently low densities 
the Bose gas enters in the so-called Tonks-Girardeau (TG) 
regime, where the system behaves as a gas of spinless 
Fermions (impenetrable Bosons) [5]. This TG regime 
has not been achieved yet, but is one of the main 
focus areas of experimental investigation. 
\par 
In this article we develop a generalized Lieb-Liniger (GLL) 
theory to investigate the crossover from a 
3D Bose-Einstein condensate (BEC) to a 1D TG gas. 
In GLL theory the $N$-body wave function is factorized 
in a cylindric-longitudinal axial wave function and a 
cylindric-radial transverse wave function 
as recently done by Das, Girardeau and 
Wright (DGW) [6]. While DGW have numerically calculated the transverse 
wave function, in our GLL approach the transverse wave function 
has a Gaussian shape with its width as variational parameter. 
In the case of an axially uniform Bose gas the GLL theory gives 
a ground-state energy very close to the one calculated in [6]. 
We use the GLL theory, which is computationally much less demanding 
than DGW method, to study dynamical properties of 
nonuniform gases extending our previous analysis of the 3D$\to$1D 
transition with BECs [7] by taking into account 
1D quantum correlations and also the effect 
of an attractive interaction.  

\section{Bose gas under transverse confinement} 

The Hamiltonian of a gas of $N$ interacting identical 
Bose atoms confined in the transverse cylindric radial direction 
with a harmonic potential of frequency $\omega_{\bot}$ 
is given by 
\beq
{\hat H} = \sum_{i=1}^N \left( -{1\over 2} \nabla_i^2
+ {1\over 2}(x_i^2+y_i^2) \right) 
+ \sum_{i<j=1}^N V({\bf r}_i,{\bf r}_j) \; , 
\eeq 
where $V({\bf r}_i,{\bf r}_j)$ is the inter-atomic interaction. 
In the Hamiltonian we have used scaled units: 
energies are in units of the energy 
$\hbar \omega_{\bot}$ of the transverse confinement 
and lengths in units of the characteristic harmonic 
length $a_{\bot}=\sqrt{\hbar/(m\omega_{\bot})}$. 
We set $V({\bf r}_i,{\bf r}_j)= g_{3D} \; 
\delta^{(3)}({\bf r}_i-{\bf r}_j)$, 
where $g_{3D}= 4\pi a_s/a_{\bot}$ with $a_s$ 
the s-wave scattering length of the inter-atomic potential. 
This Fermi pseudo-potential gives the correct dilute gas limit 
and a well-posed variational problem by choosing a smooth trial 
wave function [6]. Note that a confinement-induced resonance (CIR) 
has been predicted at $a_s/a_{\bot}\simeq 1$ [8], i.e.  
when the absolute value $|E_B|$ of 
the highest bound-state energy $E_B \simeq -\hbar^2/(ma_s^2)$ 
of a realistic inter-atomic potential approaches the confining 
transverse energy $\hbar \omega_{\bot}=
\hbar^2/(ma_{\bot}^2)$ [9]. Therefore our analysis will be 
limited to the range $|a_s|/a_{\bot} \lesssim 10^{-1}$, 
a regime where CIR effects are negligible [10]. 
\par 
The determination of the $N$-body wave function 
$\psi({\bf r}_1,...,{\bf r}_N)$ that minimizes the energy 
${\langle \psi |{\hat H}| \psi \rangle}$ of the system 
is a very difficult task. Nevertheless, 
due to the symmetry of the problem, a variational trial wave 
function can be written in the form 
\beq 
\psi({\bf r}_1,...,{\bf r}_N) = 
f(z_1,...,z_N) 
\prod_{i=1}^N { 
\exp{ \left(-{x_i^2+y_i^2\over 2 \sigma^2} \right) } 
\over \sqrt{\pi} \sigma} \; , 
\eeq 
where $\sigma$ is a variational parameter that gives the effective 
transverse length of the Bose gas. 
For repulsive inter-atomic interaction 
one has $\sigma \geq 1$ and only in the case of very strong 
transverse confinement $\sigma =1$ (1D system). 
\par 
By using Eq. (2) and integrating 
over $x$ and $y$, the total energy per unit 
of length reads ${\cal E} = {\cal E}_z + {\cal E}_{\bot}$, 
where the longitudinal axial energy is 
$
{\cal E}_z = \langle f |
\sum_{i=1}^N  -{1\over 2} {\partial^2 \over \partial z_i^2}  
+ g_{1D}(\sigma) \sum_{i<j=1}^N \delta(z_i - z_j) 
| f \rangle /L \; , 
$
with $g_{1D}(\sigma ) = g_{3D}/(2\pi \sigma^2)$ 
the effective one-dimensional inter-atomic strength, and 
the transverse radial energy is 
${\cal E}_{\bot} = {1\over 2} 
({1\over \sigma^2}+\sigma^2) \rho$, 
with $\rho=N/L$ is the homogeneous linear density 
of the Bose gas.
\par 
Due to the Lieb-Liniger theorem [3], for $a_s>0$ the axial energy 
per unit length ${\cal E}_z$ can be written in terms of the 
axial density $\rho$ only and the total energy 
${\cal E}$ can thus be rewritten as 
\beq 
{\cal E} = {1\over 2} \rho^3 
e({\gamma \over \rho \sigma^2}) 
+ {1\over 2} ({1\over \sigma^2}+\sigma^2) \rho \; , 
\label{u0} 
\eeq 
where $\gamma=2a_s/a_{\bot}$ is the inter-atomic 
strength and $e(x)$ is the Lieb-Liniger function, 
which is defined as the solution of a Fredholm equation 
and it is such that $e(x)= x-4/(3\pi)x^{3/2}$ for $x\ll 1$ and 
$e(x) = (\pi^2/3)(x/(x+2))^2$ for $x\gg 1$ [3,11]. 

\begin{figure}
\centerline{\psfig{file=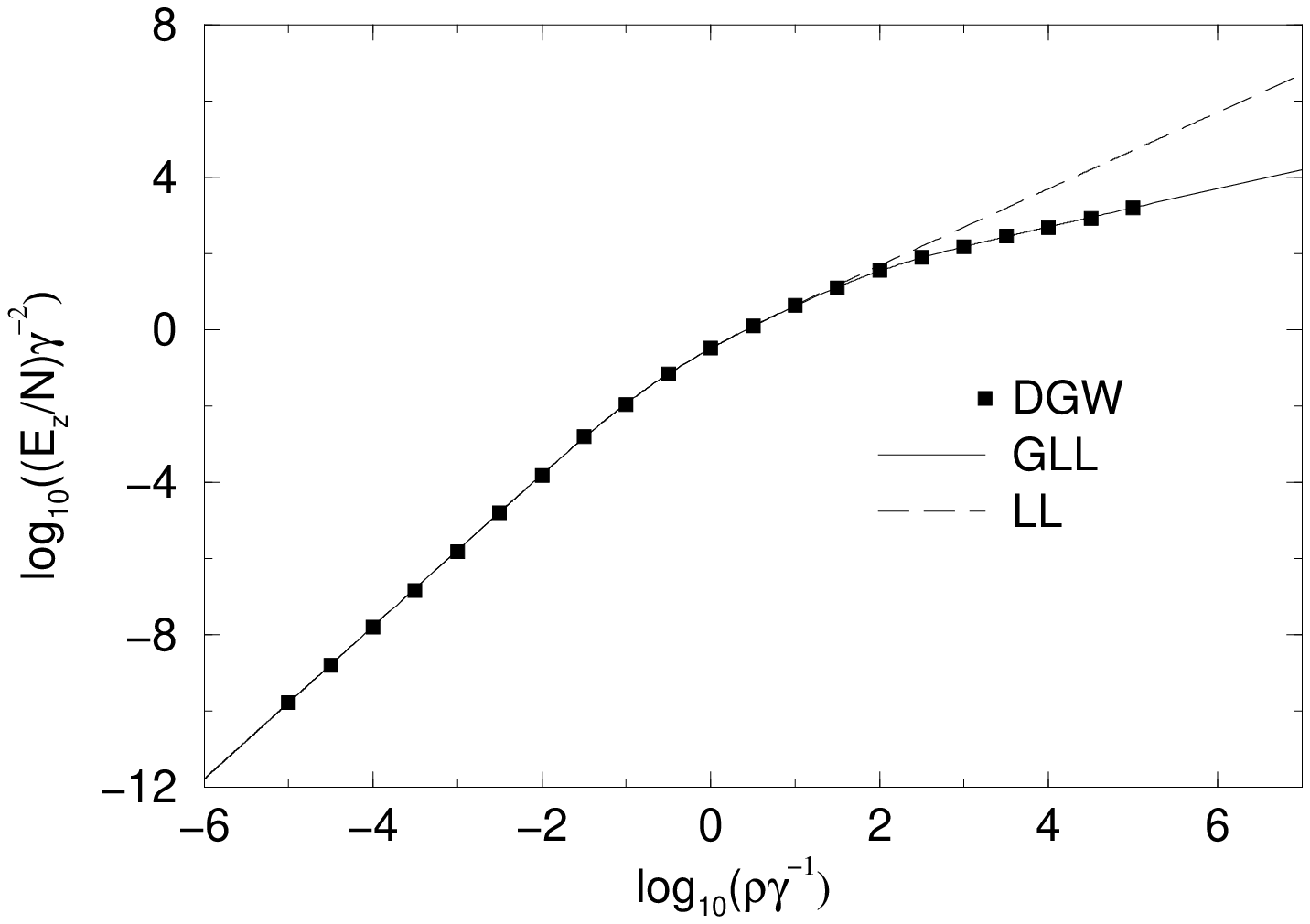,height=2.3in}}
\small 
{Fig. 1. Axial energy per particle $E_z/N$ versus 
axial density $\rho$ for a Bose gas under transverse 
harmonic confinement of frequency $\omega_{\bot}$ 
with $\gamma =2a_s/a_{\bot}=0.1$. Comparison between 
Lieb-Liniger (LL), generalized Lieb-Liniger (GLL) 
and Das-Girardeau-Wright (DGW) theories. 
The points of DGW theory are taken from Fig. 1 of paper [6]. 
Energy in units $\hbar\omega_{\bot}$, length in units $a_{\bot}$, 
and axial density in units $1/a_{\bot}$. } 
\end{figure}

\par 
The energy $\cal E$ depends on a variational parameter: 
the transverse width $\sigma$. The minimization of the energy 
${\cal E}$ with respect to $\sigma$ gives the equation 
\beq 
\sigma^4 = 1 + \gamma \rho e'({\gamma \over \rho \sigma^2}) \; .  
\label{u1} 
\eeq 
From this nontrivial equation, for a fixed value of $\gamma$,  
one numerically finds 
the transverse width $\sigma$ of the Bose gas as a function 
of the longitudinal density $\rho$. 
Analytical results can be obtained in limiting cases. 
In particular, under the condition $\gamma < 1$, 
one gets $\sigma=\sqrt{\gamma \rho}$ for $\rho \gg 1/\gamma$ 
(3D regime) and $\sigma = 1$ for 
$\rho \ll 1/\gamma$ (1D regime). Note that 
the TG regime is reached at low densities: $\rho \ll \gamma$. 
\par 
In Figure 1 we plot the axial energy 
per particle $E_z/N$ as a function of the axial density 
$\rho$ obtained by using different theories: the present 
generalized Lieb-Liniger (GLL), 
Lieb-Liniger (LL) [3], and Das-Girardeau-Wright (DGW) [6]. 
The results clearly show that GLL approach gives practically 
the same results of the the DGW theory, that is based on a 
fully numerical treatment of the transverse wave function [6]. 
Moreover, Figure 1 shows that the LL theory, obtained 
from Eq. (\ref{u1}) simply setting $\sigma=1$, is not reliable 
at high axial densities because the Bose gas 
becomes three-dimensional. 

\begin{figure}
\centerline{\psfig{file=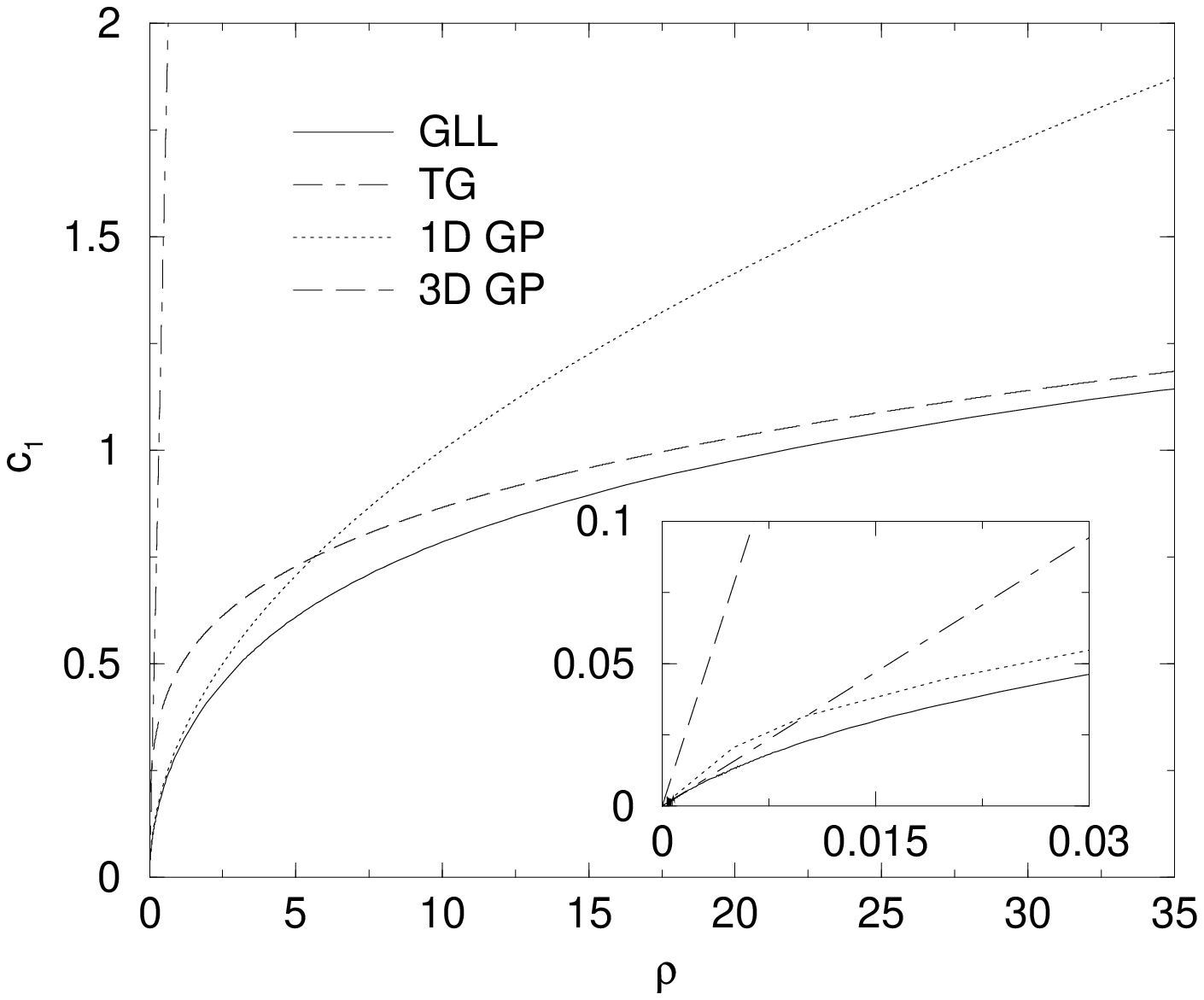,height=2.7in}}
\small 
{Fig. 2. First sound axial velocity $c_1$ versus 
axial density $\rho$ for a Bose gas under transverse 
harmonic confinement of frequency $\omega_{\bot}$ 
with $\gamma=2a_s/a_{\bot}=0.1$. Length in units $a_{\bot}$, 
time in units $\omega_{\bot}^{-1}$ and 
axial density in units $1/a_{\bot}$. } 
\end{figure} 

\par 
Having derived the function $\sigma(\rho)$, one 
can numerically determine, for a fixed $\gamma$, 
the chemical potential $\mu={\partial {\cal E}\over \partial \rho}$ 
as a function of the density $\rho$. It is given by 
\beq 
\mu = {3\over 2} \rho^2 e({\gamma \over \rho \sigma^2}) 
+ {1\over \sigma^2}  \; .  
\label{u2}
\eeq 
Again, analytical results are obtained in limiting cases. In particular, 
under the condition $\gamma < 1$, 
one finds $\mu = \sqrt{2}(\gamma \rho)^{1/2}$ 
for $\rho \gg 1/\gamma$ (3D regime), 
$\mu = \gamma \rho+1$ for 
$\gamma \ll \rho \ll 1/\gamma$ (1D regime), and 
$\mu = 3\pi^2 \rho^2/6 +1$ for $\rho \ll \gamma$ (TG regime). 
\par 
The chemical potential $\mu$ can be used to derive 
the axial sound velocity $c_1$ of the Bose gas,  
that is obtained with the formula $c_1 = 
\sqrt{\rho {\partial \mu\over \partial \rho}}$ [13]. 
In Figure 2 we plot the sound velocity $c_1$ as a function 
of the density $\rho$ for different values of the 
inter-atomic strength $\gamma$. The results show that 
our generalized Lieb-Liniger (GLL) theory works 
remarkably well. It gives the 3D BEC result of GP theory 
at high densities and the TG result of LL theory at low densities. 

\section{Including a longitudinal external field} 

Now we consider the effect of a longitudinal 
confinement due to an external potential $U(z)$. 
Assuming that the density $\rho(z,t)$ varies 
sufficiently slowly, i.e. the local density approximation (LDA), 
at each point $z$ the gas can be considered in local 
equilibrium, and the local chemical potential 
$\mu[\rho (z,t)]$ is provided by Eq. (\ref{u2}) supplemented 
by Eq. (\ref{u1}). 
Under LDA the dynamics can be described by means of 
a complex classical field $\Phi(z,t)$ that satisfies 
the time-dependent GLL nonlinear Schr\"odinger equation 
\beq 
i {\partial \over \partial t} \Phi = 
\left( - {1\over 2}{\partial^2 \over \partial z^2} 
+ U(z) + \mu[\rho] \right) \Phi \; , 
\label{f1}
\eeq
where $\rho(z,t)=|\Phi(z,t)|^2$. 
This equation is a Euler-Lagrange equation obtained by 
imposing the least-action principle to the following GLL action 
functional 
\beq 
A = \int \left\{ 
\Phi^* \left( i {\partial \over \partial t}
+{1\over 2}{\partial^2 \over \partial z^2} - U(z) \right) \Phi 
- {\cal E}[\rho , \sigma ] \; 
\right\} dz \; ,  
\label{f2} 
\eeq 
where ${\cal E}$ is given by Eq. (3), and minimizing with respect 
to $\Phi^*$. The other Euler-Lagrange 
equation is precisely Eq. (\ref{u1}), that can be obtained 
from $A$ by minimizing with respect to $\sigma$. 
It is important to observe that at high 
densities ($\rho \gg 1/\gamma$) 
$A$ gives the action functional of the non-polynomial 
Schr\"odinger equation (NPSE) we have derived to 
describe BECs under transverse confinement [7], while 
at low densities ($\rho \ll 1/\gamma$) and 
in the static case $A$ reduces 
exactly to the energy functional recently introduced by 
Lieb, Seiringer and Yngvason [4]. 
Note that Eq. (\ref{f1}) is equivalent 
to the two hydrodynamics equations of a viscousless 
fluid in 1D with density field $\rho(z,t)$ and 
velocity field $v(z,t) = - i {\partial ln(\Phi/|\Phi|) \over \partial z}$ 
(see for instance [12]). 
\par 
Neglecting the kinetic gradient term 
${1\over 2}{\partial^2 \Phi\over \partial z^2}$ 
(Thomas-Fermi approximation) 
and imposing the equilibrium condition, 
we obtain the equation 
\beq 
\mu[\rho_0(z)] = \mu_T - U(z) \; , 
\label{local} 
\eeq 
where $\rho_0(z)$ is the equilibrium density and $\mu_T$ 
is the chemical potential of the axially inhomogeneous system. 
Inverting Eq. (\ref{local}) one obtains 
$\rho_0(z)= \mu^{-1}[\mu_T - U(z)]$. 
The value of $\mu_T$ is found by imposing the normalization 
condition $N=\int \rho_0(z) dz$. 
In the case of axial harmonic confinement with scaled potential 
$U(z)={1\over 2} \lambda^2 z^2$, where 
$\lambda = \omega_z/\omega_{\bot}$ is the anisotropy 
of the full harmonic trap with $\omega_z$ axial harmonic 
frequency, the normalization condition 
can be rewritten in the form 
\beq 
\int_0^{1}\mu^{-1}[\mu_T(1-\zeta^2)] d\zeta = 
{\lambda N\over 2 \sqrt{2 \mu_T} } \; ,  
\label{pippo}
\eeq 
where $\zeta = z\lambda/\sqrt{2\mu_T}$ is the rescaled 
axial coordinate. 

\begin{figure}
\centerline{\psfig{file=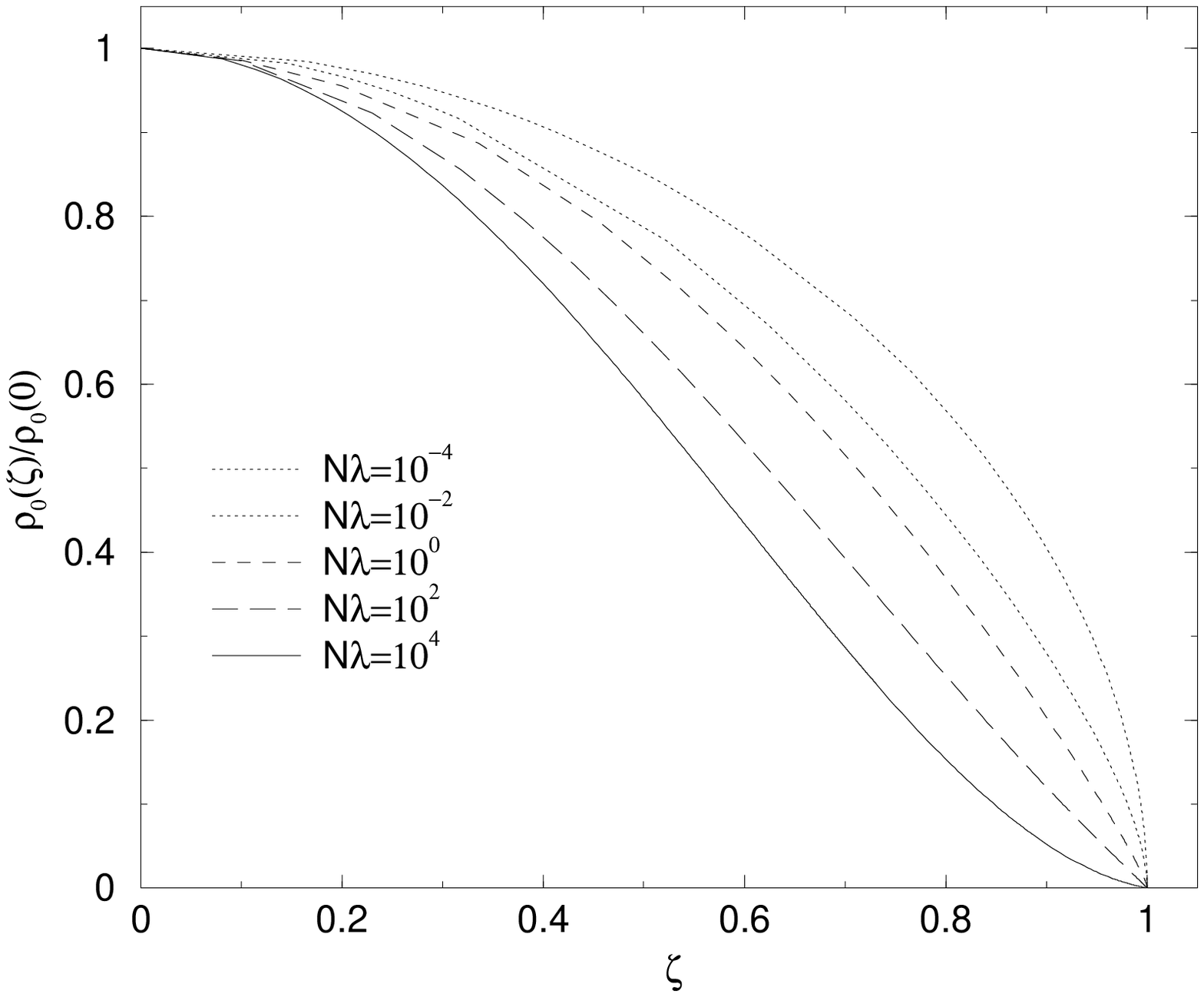,height=2.6in}}
\small
{Fig. 3. Scaled axial density profile $\rho_0(\zeta)/\rho_0(0)$ 
versus scaled axial coordinate $\zeta=z\lambda/\sqrt{2\mu_T}$ 
for a Bose gas of $N$ atoms in a harmonic confinement 
with $\gamma =2a_s/a_{\bot}= 0.1$. 
$\lambda =\omega_z/\omega_{\bot}$ is the trap anisotropy 
and $\mu_T$ is the chemical potential. Units as in Fig. 1.} 
\end{figure} 

In Figure 3 we plot the axial density $\rho_0(\zeta )$. In order 
to compare different cases, we rescale the density with respect 
to the central one $\rho_0(0)$. We have verified that 
$\rho_0(\zeta )$ follows with good accuracy the law 
$\rho_0(\zeta) = \rho_0(0)(1-\zeta^2)^{\alpha}$, where the 
exponent $\alpha$ ranges between $\alpha = 2$ (TG regime) 
to $\alpha=1/2$ (3D BEC regime) by increasing $N\lambda$.  
\par 
The collective oscillations are determined by writing the density field 
in the form $\rho(z,t)=\rho_0(z)+e^{-i\Omega t} \rho_1(z)$, 
with the function $\rho_1(z)$ obeying a linearized 
equation which follows from hydrodynamic equations [12,13]. 

\begin{figure}
\centerline{\psfig{file=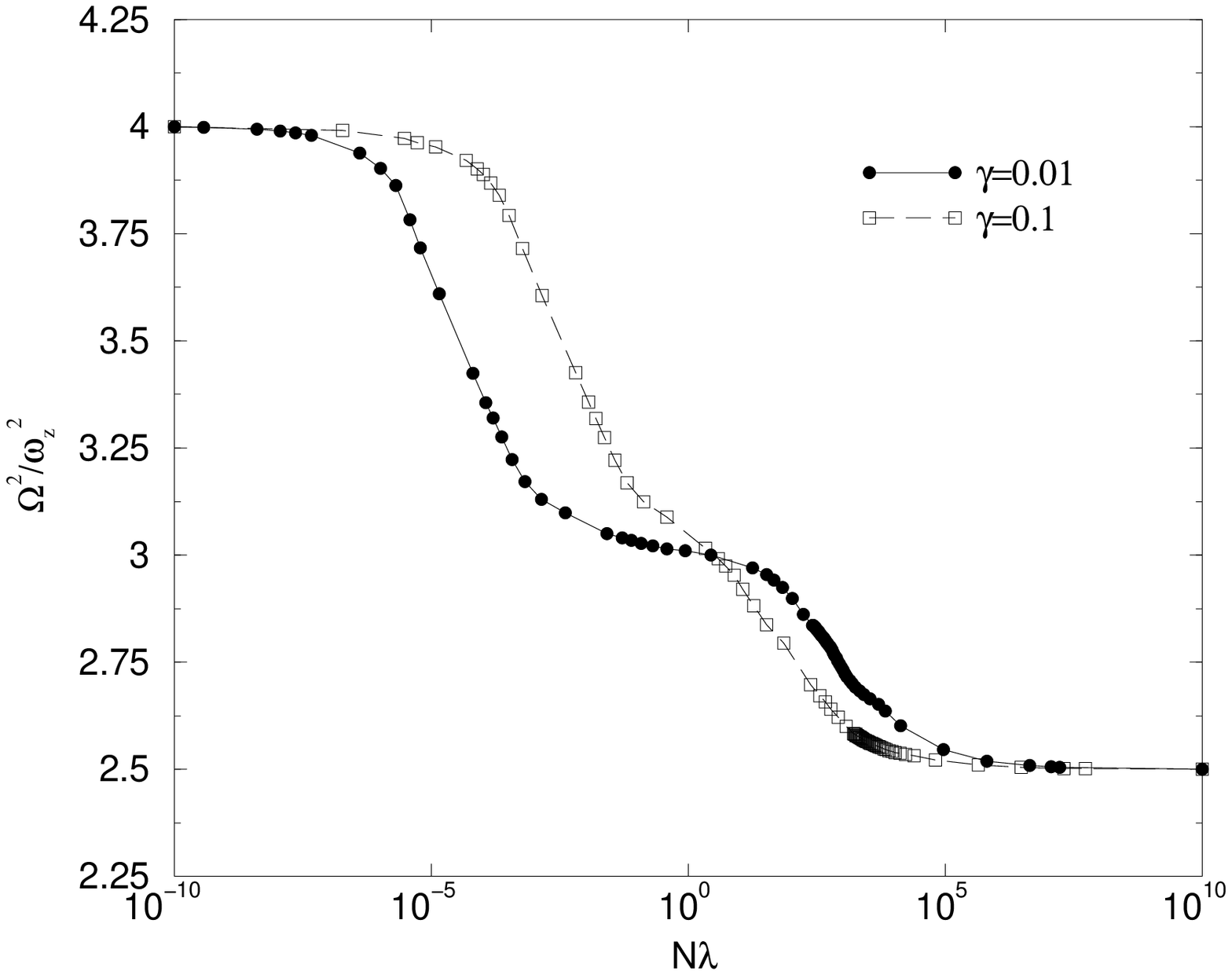,height=2.6in}}
\small 
{Fig. 4. Frequency $\Omega$ of the lowest 
compressional mode of a Bose gas of $N$ atoms in 
a harmonic confinement. 
$\lambda =\omega_z/\omega_{\bot}$ is the trap anisotropy 
and $\gamma=2a_s/a_{\bot}$ is the 
inter-atomic strength. Units as in Fig. 2. } 
\end{figure} 

The problem has simple analytic solutions 
if the equilibrium density $\rho_0(z)$ has the form 
$\rho_0(z)=\rho_0(0)(1-z^2/Z^2)^{\alpha}$. 
In this case the lowest axial compressional mode is given by 
\beq 
{\Omega^2\over \omega_z^2} = 2+ {1\over \alpha} \; . 
\eeq 
For a fixed value of $\gamma$ we easily determine $\alpha$ as 
a function of $N\lambda$ using Eq. (\ref{u1}), Eq. (\ref{u2}), 
Eq. (\ref{local}) and Eq. (\ref{pippo}). 
In such a way we obtain the lowest axial compressional mode 
$\Omega$ as a function of $N\lambda$, as shown in Figure 4. 
As expected, while for a large $\gamma$ there are two plateaus 
corresponding to $\Omega^2/\omega_z^2 =5/2$ at high densities (3D GP) 
and $\Omega^2/\omega_z^2=4$ at low densities (TG), for $\lambda \ll 1$ 
an additional plateau appears with 
$\Omega^2/\omega_z^2 = 3$ at intermediate densities (1D GP). 
\par
The results shown in Figure 4 improve 
those obtained by Menotti and Stringari [12] 
using a sum-rule approach in both 
LL theory (1D regime) and GP theory (3D BEC regime). 
In Ref. [12] they found that the curves of 
$\Omega^2/\omega_z^2$ obtained with the two theories 
do not match at intermediate densities 
if $\gamma \simeq 0.1$. In fact, LL theory predicts 
the achievement of the 1D mean-field regime 
when GP theory already exhibits 
significant 3D effects. As clearly shown in Figure 4, 
GLL theory overcomes such difficulties. 
\par 
Here we have shown that the time-dependent LDA reproduces 
known results in the appropriate limits and therefore 
can be safely used to calculate collective oscillations 
(one-phonon excitations) around static configurations of the Bose gas.  
Note that, in general, time-dependent LDA can give incorrect 
results for more complex dynamical 
phenomena, like multi-phonon excitations [14] or 
the interference of two Bosonic clouds [15]. 

\section{Negative scattering length} 

Our variational approach can be applied also 
to the case of negative 3D scattering length ($a_s<0$) 
by using Eq. (\ref{f1}) and Eq. (\ref{f2}) and simply 
setting $e(x)=x$ with $x<0$ in Eq. (\ref{u0}), Eq. (\ref{u1}) and 
Eq. (\ref{u2}). This corresponds to the Hartree 
approximation $\prod_{j=1}^N \phi(z_j)$ 
of the the exact N-body wave function $f(z_1,...,z_N)$ [7]. 
In the pure 1D regime ($\sigma=1$), Calogero and Degasperis [16] 
have shown that with negative scattering length 
the exact N-body wave function 
is well approximated by the Hartree ansatz. 
In this way Eq. (\ref{f1}) reduces to the NPSE,  
Eq. (\ref{u1}) gives the equation 
$\sigma^2=\sqrt{1 - |\gamma|\rho }$ for the 
transverse width [17]. 
\par 
In a toroidal configuration, i.e. cylindric confinement 
with periodic boundary conditions in the axial 
direction and $U(z)=0$, the Bose gas with negative 
scattering length is axially uniform below a critical 
number $N_c$ of Bosons and becomes localized (bright soliton) 
only above this critical number. 
Carr, Clark and Reinhardt [18] have shown that 
in the pure 1D regime ($\sigma =1$) this critical number is given by 
$N_c= \pi^2/(|\gamma|L)$, where $L$ is the length 
of the torus. We easily extend 
that prediction to a 3D Bose gas by calculating with NPSE 
the wave-length $\lambda_c$ of the zero-energy Bogoliubov 
excitation above the uniform state [19]: 
in this way, setting $\lambda_c = L$, 
we find the equation $ N_c|\gamma|L/\pi^2 = (1-N_c|\gamma|/L)^{3/2}/
(1-3N_c|\gamma|/(4L))$. Contrary to the 1D theory, 
our GLL theory predicts the existence of the single bright soliton 
in toroidal confinement only below a critical strength $N|\gamma|$. 

\begin{figure}
\centerline{\psfig{file=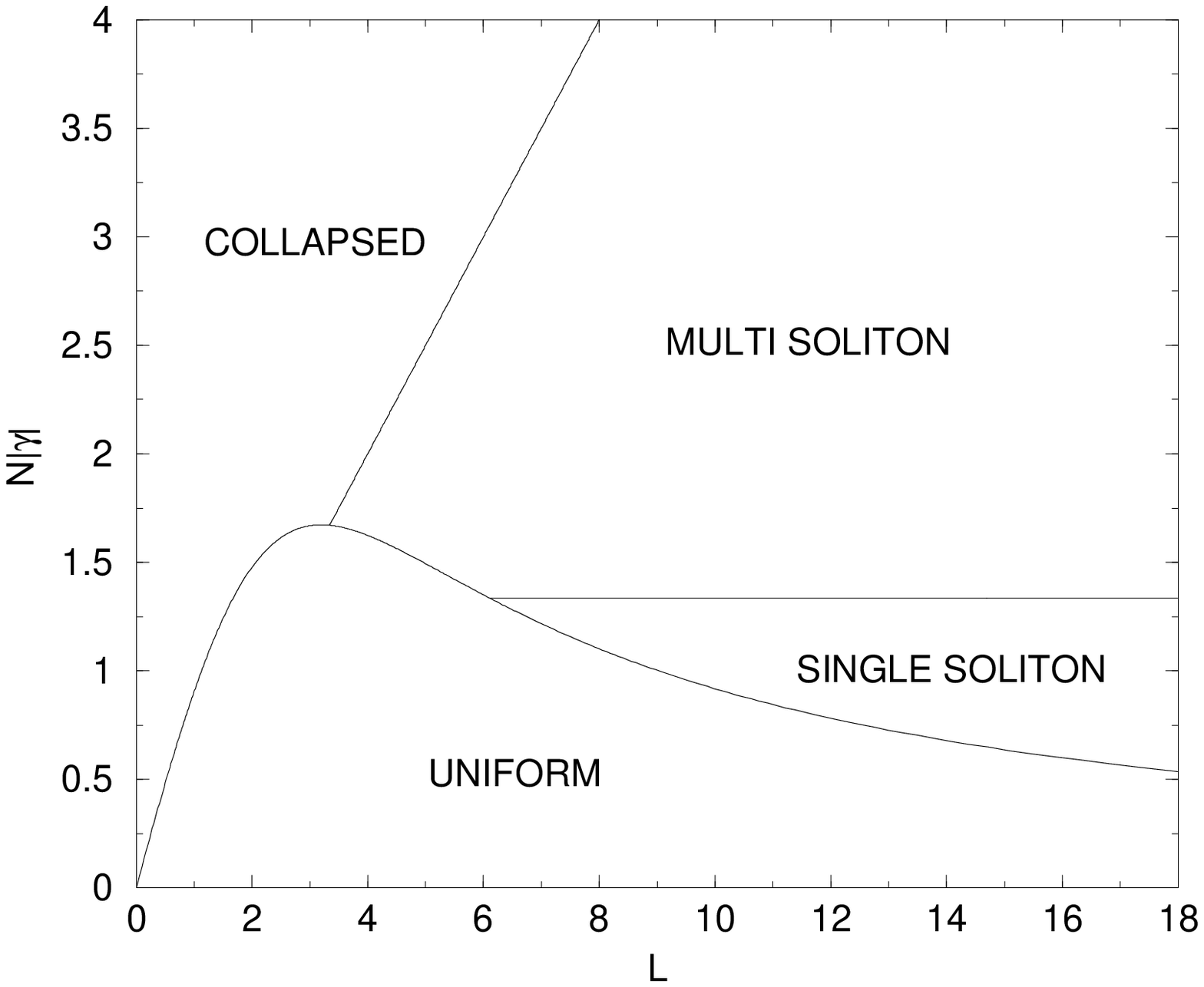,height=2.6in}}
\small 
{Fig. 5. Phase diagram of a Bose gas with negative 
scattering length in toroidal confinement. 
$N$ is the number of Bosons, 
$\gamma=2a_s/a_{\bot}$ is the inter-atomic strength 
and $L$ is the length of the torus. Units as in Fig. 1.} 
\end{figure} 

We have numerically verified that such a critical strength is 
well represented by $N|\gamma| = 4/3$. 
Above this line only multi-soliton configurations are possible, 
which collapse at $N|\gamma| = L/2$ as in the case of the NPSE 
with box boundary conditions (see [17]). 
In Figure 5 we show the phase diagram of the toroidal Bose gas 
with negative scattering length in the plane 
$N|\gamma|$ {\it vs} $L$: uniform, localized with 
a single bright soliton, localized with a 
multi soliton train and collapsed. 
The properties of these multi-soliton configurations 
will be discussed elsewhere. 

\section{Conclusions} 

We have introduced a very effective generalized 
Lieb-Liniger theory, based on a variational treatment of 
the transverse width, that points out the crucial interplay between 
quantum correlations and dimensionality in confined Bose systems. 
The first sound velocity, the density profile and the 
collective oscillations give a clear 
signature of the regime involved. 
Our results exactly reproduces the ones obtained in specific 
regimes with the hydrodynamic method or the sum-rule method, but, 
they also give the crossover from one regime to another. 
Finally, we have applied our approach to a Bose gas with negative 
scattering length deriving the complex phase diagram 
(that includes uniform, single-soliton and multi-soliton 
configurations) of an attractive Bose gas in toroidal confinement.


\begin{thebibliography}{99}
\bibitem{11} $e(x)$ and its derivative $e'(x)$ are  
obtained numerically with an iterative algorithm.
\end{thebibliography}
\end{document}